\documentclass{PoS}
\usepackage{epsfig}
\usepackage{psfrag}

\newcommand{\al} {\alpha}

\title{Lattice String Field Theory}

\ShortTitle{Lattice String Field Theory}

\author{\speaker{Francis Bursa}\\
        Jesus College, University of Cambridge\\
        E-mail: \email{fwb22@cam.ac.uk}}

\author{Michael Kroyter\\
        CTP, Massachusetts Institute of Technology and Tel-Aviv University\\
        E-mail: \email{mikroyt@tau.ac.il}}

\abstract{String field theory is a candidate for a full non-perturbative
definition of string theory. We aim to define string field theory on
a space-time lattice to investigate its behaviour at the quantum
level. Specifically, we look at string field theory in a one dimensional
linear dilaton background. We report the first results of our simulations.}

\FullConference{The XXVIII International Symposium on Lattice Field Theory, Lattice2010\\
		June 14-19, 2010\\
		Villasimius, Italy}

\begin{document}

\section{Introduction}

Superstring theory is currently the most promising candidate for
a theory of everything. Yet, it is not clear what superstring theory is.
Moreover, even the perturbation theory of the standard (RNS)
formulation of string theory is not yet completely established beyond some
loop level, due to complications related to supermoduli
spaces~\cite{D'Hoker:2002gw,Grushevsky:2008zm,DuninBarkowski:2009ej}.
A possible way to define superstring theory that might also resolve the
problems with supermoduli spaces, is as a superstring field theory, i.e.,
as a field theory of strings (see~\cite{Fuchs:2008cc} for a recent review).
It is natural to expect from a reliable formulation of superstring field
theory that it respects the underlying symmetries
of string theory, i.e., it should be covariant and, moreover,
universal~\cite{Sen:1999xm}.
There are several variants of superstring field theories of this kind, e.g.,
the heterotic~\cite{Berkovits:2004xh} and
open~\cite{Arefeva:1989cp,Preitschopf:1989fc,Berkovits:1995ab,Kroyter:2009rn}
theories. However, it was never checked if any of those is really well
defined at the quantum level. It is our intention to address this question
using lattice techniques.

Some of the formulations of superstring field theory cannot be expected, at
this stage, to be consistent at the quantum level, since, e.g., they don't
include a consistent Ramond sector. The fermions of the Ramond sector, as
well as the notion of ``picture''~\cite{Friedan:1985ge} introduce, in any
case, some complications for the theory.
Hence, it might be advisable to start with a simpler model,
such as that of the bosonic string. The bosonic closed string field
theory~\cite{Zwiebach:1992ie} is much more complicated than the open
one~\cite{Witten:1986cc}. Hence, we concentrate on the later.

It is known that the bosonic string theory is consistent in flat space only
in 26 dimensions. Simulating any theory on a 26 dimensional lattice is almost
a hopeless task. Moreover, the theory has tachyons, both in the open and in
the closed string spectrum. It is by now understood that the open string
tachyon is related to a condensation of an unstable
D-brane~\cite{Sen:1999xm,Sen:1999mh,Sen:1999nx,Schnabl:2005gv}. However,
no analogous understanding exists regarding the fate of the closed string
tachyon~\cite{Yang:2005rw,Yang:2005rx,Moeller:2006cv}.
Both problems can be avoided by considering ``non-critical'' string theory.
The non-critical theory lives at lower dimensions and for $d\leq 2$ the
tachyon is absent. On the other hand, a new complication is introduced,
namely the theory includes a linear dilaton, which breaks Poincar\'e
invariance. Moreover, all the coupling constants are proportional to the
dilaton vacuum expectation value, which runs to infinity in one direction. These issues pose
a challenge to a lattice simulation.

A simple consideration of world-sheet gauge symmetry and degrees of freedom
reveals that two of the $d$ dimensions in which the string lives are
unphysical. Indeed, the two-dimensional string theory is already not a
string theory, but a field theory, in the sense that only one physical field
out of the infinitely many ones, remains. This field is the ``tachyon'',
which is now a massless field. Going on to ``lower dimensions'' is possible.
The dimension is replaced by the central charge of the conformal field
theory. This is natural, since $d$ flat dimensions correspond to $c=d$,
where $c$ is the central charge. The two-dimensional theory includes a single
flat scalar field, i.e., a $c=1$ system, coupled to a single linear dilaton
direction. Theories with $0\leq c<1$ exist and are well studied. They go
under the name of ``minimal models''~\cite{Belavin:1984vu,9304011}. For
simplicity we are starting this study with the simplest model of all,
the $c=0$ theory that includes only the one dimensional linear dilaton
direction with $c=26$ and the canonical ghost system with $c=-26$.

While minimal models have only a finite number of degrees of freedom,
the string field theory that describe them still contains an infinite
number of fields. Almost all of the degrees of freedom of the theory should
then be removed by the very large gauge symmetry present. This fact raises
two further problem that one should address. First, we have to truncate the
infinite number of fields to a finite number while taking this number to
infinity eventually. Second, we have to address the existence of the gauge
symmetry. The first issue of ``level truncation'' was much employed in the
string field theory
literature~\cite{Kostelecky:1990nt,Kostelecky:1988ta,Gaiotto:2002wy},
albeit only at the classical level. It is
not even a-priori clear that it would be a consistent regularization at
the quantum level. Here, we apply an ``experimental approach'' towards this
question. The issue of gauge symmetry arises only at the next level.
Since in this report we only concentrate on preliminary results from the
lowest level, we ignore this issue for now. 

\section{Methods}

The ``level'' of level truncation is, up to an additive constant, the
eigenvalue of the ``Hamiltonian'' $L_0$ (the zeroth Virasoro generator).
As such, the level includes two contributions,
that of the field itself, which is different for any of the infinitely many
fields of which the string field is composed and a momentum contribution for
each possible mode of the field. The former contribution is denoted $l_0$ and
the total level is given by,
\begin{equation}
l=l_0+\al' p^2\,,
\end{equation}
where $p$ is the momentum and $\al'$ is a dimensional constant setting the
string scale. We also assumed that the fields were properly redefined,
e.g., instead of the canonical ``tachyon field'' $T(x)$ we consider
$\tau(x)=e^{-\frac{Vx}{2}}T(x)$. This is the only field with $l_0=0$.

The $l_0=0$ level action is
\begin{equation}
S=-\frac{1}{2}\int d x\, \big(m_0^2\tau^2+(\nabla \tau)^2\big)
 -\frac{g_o K^{3\big(1-\frac{\al' V^2}{4}\big)}}{3}\int d x\,
 e^{-\frac{V \cdot x}{2}}\tilde \tau (x)^3 \,,
\end{equation}
where $K=\frac{3\sqrt{3}}{4}$, $g_o$ is the open string coupling constant,
$V$ is the dilaton gradient
$V=-\sqrt{\frac{25}{6\al'}}$ and $m_0^2$ is the mass squared of the
``tachyon field'', $m_0^2=\frac{V^2}{4}-\frac{1}{\al'}=\frac{1}{24\al'}$.
The second term depends on a non-local variant of $\tau$, namely
$\tilde \tau(x)=K^{\al' \nabla^2} \tau(x)$.
This action is both non-local and space-dependent. Furthermore, we
cannot use periodic boundary conditions, since this would unphysically
glue together the strong-coupling region at large $x$ with the
weak-coupling region at small $x$. Instead we choose Dirichlet
boundary conditions and expand $\tau(x)$ on an interval $x_{min}< x <
x_{max}=x_{min}+L$ in sine waves:
\begin{equation}
\tau(x)=\sqrt{\frac{2}{L}}\sum_{n=1}^N
  \tau_n \sin\Big(\frac{\pi n x}{L}\Big)\,.
\end{equation}
Here the level of each mode is given by $l(\tau_n)=\al' (\frac{\pi
  n}{L})^2$. We choose $N$ so that all modes have $l < 1$ since we are
working at zero $l_0$ level. We also set $g_o=1$, which amounts to a shift
in $x$.

In terms of the $\tau_n$, the action is
\begin{equation}
S=-\frac{1}{2}\sum_{n=1}^N \Big(\frac{1}{24\al'}
  +\big(\frac{\pi n}{L}\big)^2\Big) \tau_n^2
-\frac{g_o K^{3\big(1-\frac{\al' V^2}{4}\big)}}{3} \sum_{n_{1,2,3}=1}^N
   K^{-\al'\big(\frac{\pi}{L}\big)^2(n_1^2+n_2^2+n_3^2)}
     \tau_{n_1}\tau_{n_2}\tau_{n_3}f_{n_1,n_2,n_3}\,,
\label{tau_n action}
\end{equation}
where
\begin{equation}
f_{n_1,n_2,n_3} = \Big(\frac{2}{L}\Big)^\frac{3}{2}\int_{x_{min}}^{x_{max}} dx
  \,e^{-\frac{V x}{2}}\sin\Big(\frac{\pi n_1 x}{L}\Big)
    \sin\Big(\frac{\pi n_2 x}{L}\Big)\sin\Big(\frac{\pi n_3
      x}{L}\Big).
\label{f_n1n2n3}
\end{equation}
This is the action we want to consider. Note that the weight of a
configuration in the path integral is $e^S$ rather than $e^{-S}$ due
to the way we Wick-rotated.

We see an immediate problem: the action~(\ref{tau_n action}) has a
cubic instability. To proceed, we consider the integral $\int d
\tau_n$ over each mode as a complex integral, and deform the
integration contour to be a straight line at an angle $\gamma$ to the
real axis.
If we choose $\gamma=\pi/6$, the cubic part of the action becomes pure
imaginary and so the action is no longer unstable. This is similar to
the contour deformation used to consider an analytic continuation of
Chern-Simons theory~\cite{Witten:2010cx}.

However, taking the $\tau_n$ to be complex introduces another problem;
the action also becomes complex and so cannot be interpreted as a
weight for a Markov chain. Instead we simulate in the phase-quenched
ensemble and reweight. That is, we split $e^{S}$ into an amplitude and
a phase:
\begin{equation}
e^{S}=|e^{S}| e^{i\theta},
\end{equation}
and calculate the expectation value of an observable $\mathcal{O}$
using the identity
\begin{eqnarray}
\langle \mathcal{O} \rangle & = & \frac{\int \mathcal{O}|e^{S}|e^{i \theta}}{\int
  |e^{S}|e^{i \theta}} \\
& = & \frac{\langle \mathcal{O}e^{i \theta} \rangle_\mathrm{PQ}}{\langle e^{i
    \theta} \rangle_\mathrm{PQ}},
\end{eqnarray}
where the label $\mathrm{PQ}$ means the expectation value is evaluated in
  the phase-quenched ensemble, i.e. with the weight $|e^{S}|$. This
  is a real, positive weight, so can be used in a Monte Carlo simulation.

We generate configurations in the phase-quenched ensemble using a
Metropolis algorithm. The observables we measure are the action $S$
and the fields $\tau_n$. We estimate errors with the jackknife method.

\section{Results}
In principle, we expect that the theory will be unstable for all
values of the parameters, since there is always a cubic term in the
action. However, due to the factor $e^{-\frac{V x}{2}}$ in
$f_{n_1,n_2,n_3}$, this term can be exponentially small, in which case
the theory will be stable for all practical purposes. We have
performed scans in parameter space to search for the onset of instability.

The instability can be seen by looking at the imaginary parts of
$\langle S \rangle$ and $\langle \tau_n \rangle$, which will be zero
for a stable set of parameters and will become non-zero as the
instability increases. In practice we find that the errors are smaller
for the $\langle \tau_n \rangle$ than for the action, so we will
concentrate on the former from now on; however the behaviour of the
action is very similar.

\begin{figure}
  \centering
  \psfrag{Re<tau>}{$\mathrm{Re}\langle \tau_n \rangle$}
  \psfrag{Im<tau>}{$\mathrm{Im}\langle \tau_n \rangle$}
  \epsfig{file=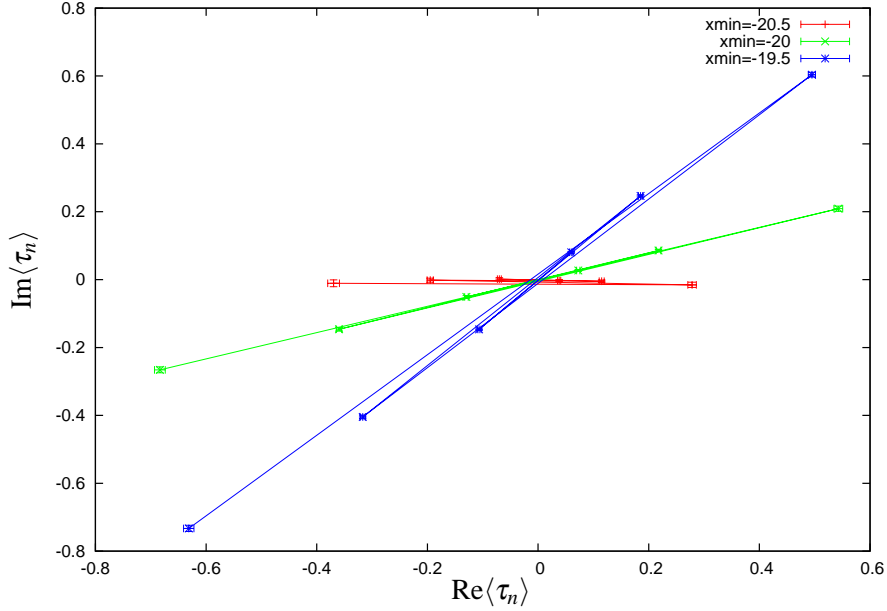,scale=0.92,clip}  
  \caption{$\langle \tau_n \rangle$ in the complex plane for
    $x_{min}=-20.5$ (red),
    $-20$ (green), and $-19.5$ (blue). All for $L=20$, $N=6$, $\al'=1$,
    $V=-\sqrt{\frac{25}{6\al'}}$. In each case $\langle \tau_1
    \rangle$ is the point furthest to the left.}
  \label{complx_fig}
\end{figure}

We find that the onset of the instability is rather rapid. We show an
example in Fig.~\ref{complx_fig}. Here we have $N=6$, which is the
maximum allowed for $L=20$. In each case we observe that the $\langle
\tau_n \rangle$ oscillate, but at $x_{min}=-20.5$ they have negligible
imaginary parts, whereas
by $x_{min}=-19.5$ the imaginary parts are as large as the real parts.
Hence the instability appears roughly when $x_{max}=0$, that is when
we start to include the region $x>0$ where the cubic terms become
large. The behaviour for other values of $L$ is very similar, with the
instability first appearing around $x_{max}=0$ in each case. 

Since the action is space-dependent, the meaning of the
infinite-volume limit is unclear. It is straightforward to decrease $x_{min}$,
since the cubic term becomes exponentially small at small
$x$. However, it is not clear if the limit $x_{max} \rightarrow
\infty$ is well-defined, since the cubic term continues to get
stronger in this direction. Indeed, we find that the imaginary parts
of the $\tau_n$ continue to increase rapidly when we increase $x_{max}$.

\subsection{Continuum limit}
\begin{figure}
  \centering
  \psfrag{l}{$l$}
  \psfrag{Im<tau1>}{$\mathrm{Im}\langle \tau_1 \rangle$}
  \epsfig{file=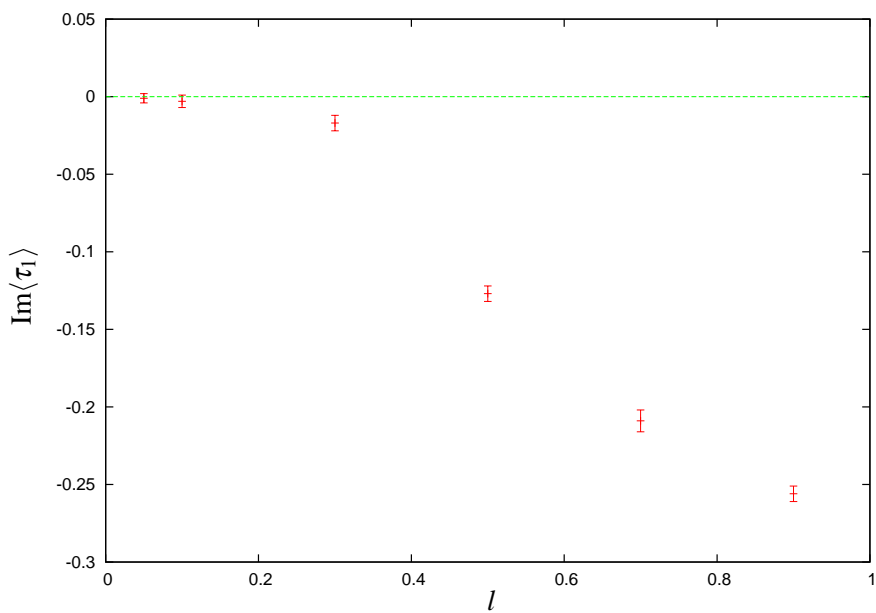,scale=0.46,angle=-90,clip}  
  \caption{$Im \langle \tau_1 \rangle$ as a function of level $l$, for
    $x_{min}=-20$, $L=20$, $\al'=1$,
    $V=-\sqrt{\frac{25}{6\al'}}$.}
  \label{contlimit_fig}
\end{figure}

In momentum space, the continuum limit is approached by increasing the
number of modes~$N$. We can only increase $N$ up to a
maximum value of $L/\pi\sqrt{\al'}$ since we require $l<1$. In this
range we find that the instability becomes stronger as $N$ is
increased, presumably because the number of unstable cubic terms
increases rapidly with $N$. We show an example in
Fig.~\ref{contlimit_fig},
where the maximum level increases from 0.05 (where $N=1$)
to 0.95 ($N=6$).
To approach closer to the continuum limit,
we will have to include more fields at higher
$l_0$-levels. Calculations at level-1 are in progress.

\subsection{Dilaton}
Finally, we have considered what happens when we vary the dilaton
$V$. In the full string field theory we require
$V=-\sqrt{\frac{25}{6\al'}}$. We might expect that varying $V$ away
from this value would increase the instability. However, this is not
the case, at least for the level-0 theory. Increasing $V$ above
$-\sqrt{\frac{24}{6\al'}}$ changes the sign of $m_0^2$, making the
theory tachyonic and hence more unstable; on the other hand decreasing
$V$ makes $m_0^2$ larger and the cubic terms smaller and hence
decreases the instability. Our results when we vary $V$ are 
consistent with this picture. However, it should be noted that
decreasing $V$ does not remove the instability entirely, and it will
always become strong at some sufficiently large value of $x$.

\section{Conclusions}
We have implemented a Monte Carlo simulation of the 1-d linear
dilation truncated to zero $l_0$-level. We observe non-trivial
quantum effects: the classical solution to the equations of motion is
$\tau_n=0$, but we observe non-zero $\tau_n$. We also find that as
expected, the theory is unstable at large $x$, as shown by the large
imaginary parts the expectation values of the field develop.

There are several possible explanations for our result. Firstly, it
may be that the instability is a real feature of the full,
non-truncated theory. This should not be the case for the theory at hand.
Another possibility is that the instability is just an artifact of the
level-truncation. In this scenario the higher-level fields, which we
have not included, would stabilise the theory. Alternatively, it might
also be the case that level-truncation is not a consistent regularization
of the quantum theory.
Finally it is also possible that the instability represents some
fundamental problem with open string field theory as a method for
quantising string theory. This could be attributed, e.g., to the lack of
control over closed string degrees of freedom or to the somewhat
singular nature of the star product.

Calculations including level-1 fields are underway. At this level it
is not obvious how to deal with the gauge and ghosts degrees of
freedom, and there are several possible choices. It will be
interesting to see how these compare. A practical issue is that
Grassmann-odd fields will appear and will have to be dealt with.

Looking further ahead, it would be interesting to increase the number
of dimensions. Ultimately the target would be to work in ten dimensions
and to include fermionic degrees of freedom, with the aim of reaching a full
quantum, non-perturbative definition of superstring theory.

\section*{Acknowledgements}
We would like to thank Y.~Oz, L.~Rastelli and B.~Zwiebach for discussions.
The research of M.~K is supported by a Marie Curie OIF. The views presented
are those of the authors and do not necessarily reflect those of the European
Community.

\end{document}